\documentclass[smallextended]{svjour3} 

\smartqed

\usepackage{amssymb,amsmath}
\usepackage[abbrev]{amsrefs}
\usepackage{xy}
\xyoption{all}
\usepackage{graphicx}
\usepackage{hyperref}
\usepackage{enumitem}

\newcommand\V{\bigvee}
\newcommand\CC{\mathbb{C}}
\newcommand\RR{\mathbb{R}}

\newcommand\Max{\operatorname{Max}}
\newcommand\ie{i.e.}
\newcommand\eg{e.g.}
\newcommand\cf{cf.}
\newcommand\st{\mid}
\newcommand\opens{\operatorname{\mathcal{O}}}
\newcommand\topology{\operatorname{\Omega}}
\newcommand\ps[1]{\widetilde{#1}}
\newcommand\spanmap{\operatorname{span}}
\newcommand\closedsets{\operatorname{\mathsf{C}}}
\newcommand\spinup{{\uparrow}}
\newcommand\spindown{{\downarrow}}

\newcommand\inv{*}
\newcommand\basechg{\boldsymbol c}
\newcommand\af{\boldsymbol\alpha}

\newcommand\wall{\mathcal{W}}
\newcommand\supp{\operatorname{supp}}
\newcommand\pair{\operatorname{Pair}}

\begin{document}

\title{An abstract theory of physical measurements\thanks{Work funded by FCT/Portugal through project UIDB/04459/2020.}
\thanks{This version of the article has been accepted for publication, after peer 
review but is not the Version of Record and does not reflect post-acceptance 
improvements, or any corrections. The Version of Record is available online at: 
http://dx.doi.org/10.1007/s10701-021-00513-1. Use of this Accepted Version is subject to the publisher's Accepted 
Manuscript terms of use https://www.springernature.com/gp/open-research/policies/accepted-manuscript-terms.}}
\author{Pedro Resende}

\institute{Centro de An\'alise Matem\'atica, Geometria e Sistemas Din\^amicos\\
Departamento de Matem\'{a}tica, Instituto Superior T\'{e}cnico,
Universidade de Lisboa\\
Av.\ Rovisco Pais 1, 1049-001 Lisboa, Portugal\\
              \email{pmr@math.tecnico.ulisboa.pt}
}

\date{Received: date / Accepted: date}

\maketitle

\begin{abstract}
The question of what should be meant by a measurement is tackled from a mathematical perspective whose physical interpretation is that a measurement is a fundamental process via which a finite amount of classical information is produced. This translates into an algebraic and topological definition of \emph{measurement space} that caters for the distinction between quantum and classical measurements and allows a notion of observer to be derived.
\keywords{Measurement problem \and Classical information \and Observers \and Quantales \and Locales \and Observational logic}
\end{abstract}

\section{Introduction}

The notion of measurement in quantum mechanics is problematic because of the apparent need to understand it in terms of concepts like ``system'' and ``apparatus'' whose definitions are elusive and therefore provide a shaky ontological foundation for one of the pillars of modern physics.
An amusing and eloquent expression of misgivings concerning ``measurers'' was voiced by Bell~\cite{Bell90}:
\begin{quote}
``It would seem that the theory is exclusively concerned about `results of measurement', and has nothing to say about anything else. What exactly qualifies some physical
systems to play the role of `measurer'? Was the wavefunction of the world waiting to jump for thousands of millions of years until a single-celled living creature appeared? Or did it have to wait a little longer, for some better qualified system... with a PhD?''
\end{quote}

This conundrum motivates many of the realist variants of quantum theory that either try to explain or do away with the ``collapse of the wave function,'' such as Bohmian mechanics~\cite{Bohm-IandII}, objective collapse theories~\cites{Ghi-etal90,Ghi-etal03,Penrose2014}, the many-worlds interpretation~\cite{Everett,Wheeler,DeWitt}, or decoherence~\cite{Zeh70,Zeh73,decoherencebook,Zurek03}. Approaches such as these can, more or less explicitly, be regarded as reductionist attempts to explain measurements, or to outright dismiss them. An explicit anti-measurement sentiment is conveyed in another eloquent passage by Bell~\cite{Bell90}:
\begin{quote}
``The first charge against `measurement', in the fundamental axioms of quantum mechanics, is that it anchors there the shifty split of the world into `system' and `apparatus'. A second charge is that the word comes loaded with meaning from everyday life, meaning which is entirely inappropriate in the quantum context. (...) In other contexts, physicists have been able to take words from everyday language and use them as technical terms with no great harm done. (...) Would that it were so with `measurement'. But in fact the word has had such a damaging effect on the discussion, that I think it should now be banned altogether in quantum mechanics.''
\end{quote}

Rather than banning the word, I propose to bootstrap a definition of measurement independently of prior distinctions between ``system'' and ``apparatus,''  ``microscopic'' and ``macroscopic,'' ``quantum'' and ``classical,'' etc. For that I take the stance that the essence of what constitutes a measurement may be a fundamental part of reality from which ``entities'' such as observers are derived. I shall tackle this from a mathematical perspective whose physical interpretation can be loosely described as follows:
\emph{by a measurement is meant any finite physical process in the course of which a finite amount of classical information is produced}, where, intuitively, by \emph{finite classical information} is meant that which can be recorded by writing down a finite list of 0s and 1s on a notebook. 
Of course, implicit in this idealization is the assumption that classical information itself must be considered fundamental, which is in line with Wheeler's ``it from bit''~\cite{itfrombit} but also sidesteps Bell's~\cite{Bell90} questions: ``\emph{Whose} information? Information about \emph{what}?''

The main aim of this paper is to introduce mathematical definitions of measurement and observer, in terms of their algebraic, topological and logical structures.
Concretely, this will be conveyed by a definition of \emph{space} of measurements. The starting point is a sober topological space whose points are the measurements and whose open sets represent physical properties, each of which corresponding to a finite amount of classical information. Perhaps not surprisingly, there is a resemblance to ideas in theoretical computer science where open sets play the role of finitely observable properties (of computers running programs)~\cite{topologyvialogic}. In addition, algebraic operations carry notions of time and causality that make any space of measurements a stably Gelfand quantale in the sense of~\cite{GSQS}. In consequence, measurement spaces have a rich symmetric structure carried by associated pseudogroups and \'etale groupoids~\cite{SGQ}.

The resulting mathematical structure can be regarded as an ``amplitude free'' quantum theory that plays a role, with respect to actual quantum theory, which is analogous to that of point set topology in relation to measure theory. An interesting feature is that classical measurements, defined by adding two logical and information processing constraints, are automatically identified with open sets of fairly well behaved topological spaces, hence leading to the emergence of spatial structures (with classical state spaces) within measurement spaces. This provides grounds for a derived notion of classical observer, albeit not external to the systems being observed but rather embedded into the systems.

\section{Schr\"odinger's electron}\label{schrelectron}

For illustration purposes let us call \emph{Schr\"odinger's electron} to an experiment in the style of Schr\"odinger's cat, but in which the cat has been replaced (much more ethically) by an electron that travels through a Stern--Gerlach apparatus. (Consider this to be a thought experiment that sidesteps the specifics of running Stern--Gerlach measurements on electrons~\cite{SGelectrons}.) The whole apparatus, including the target that records the deflection of the electron, is inside a closed box and we cannot immediately see what happens inside. All we initially know is that the magnetic field is oriented so as to measure spin along $z$.

Then there are two options: we can keep the box closed, hence obtaining no information about spin other than that deflections must be along $z$; or we open the box and obtain more classical information, namely that there was an upwards deflection or a downwards one. In terms of the classical information that can be obtained nothing distinguishes this procedure from another, using an open box, where the deflection of the electron is immediately visible. Even though the two procedures have physically different details, one point of this paper is that they are equivalent in terms of the classical information they can produce about spin, hence defining the same \emph{abstract} spin measurement.

Let us write $\boldsymbol z$ in order to denote the (abstract) measurement just described.
Let us also write $\boldsymbol z^{\spindown}$ and $\boldsymbol z^{\spinup}$ for two other measurements that can be performed using the same equipment, but observing only the lower half or the upper half of the target: $\boldsymbol z^{\spindown}$ is the measurement that records a downwards deflection but would ignore an upwards one, and $\boldsymbol z^{\spinup}$ records upward deflections while ignoring downwards ones.

We shall write $\boldsymbol x$ for a similar measurement of spin along $x$, and $\boldsymbol x^{\spindown}$ and $\boldsymbol x^{\spinup}$ are the two measurements that observe only one of the possible deflections.
Note that $\boldsymbol x$ is different from $\boldsymbol z$ because it produces different information: even keeping the boxes closed we know the possible deflections are horizontal and vertical, respectively. So we cannot represent  measurements like $\boldsymbol z^{\spindown}$ and $\boldsymbol z^{\spinup}$ by the usual rays in the Hilbert space $\CC^2$, for in that case we would have $\boldsymbol z=\boldsymbol x=\CC^2$.

\section{Measurement topology}

\subsection{Measurements and properties}

By a \emph{measurement} will informally be meant any process via which new classical information is recorded in a physical device, such as when 0s and 1s are written on a sheet of paper, or a dial of an electronic instrument indicates a new value, or there is a change in brain synapses, etc. Any finite amount of classical information which is thus associated to a measurement will be referred to as a \emph{physical property}, and two measurements are considered the same if they cannot be distinguished by their physical properties. Note that measurements should be understood to be \emph{measurement types}, meaning that multiple runs of similar measuring processes can be performed.

The way in which I shall model this is inspired by the idea of \emph{finitely observable property}~\cite{topologyvialogic} that arose in computer science in the study of the semantics of programming languages: measurements will be taken to be the points of a topological space $M$ whose open sets represent the physical properties. The relation between measurements and properties is expressed simply by saying that if $m$ is a measurement and $U$ is a physical property, then $m\in U$ means that $U$ is \emph{compatible} with $m$. This does not mean that $U$ is necessarily recorded if $m$ is performed, only that it \emph{can} be recorded. Note that an inclusion $U\subset V$ plays the role of a \emph{logical implication}: being compatible with $U$ implies being compatible with $V$.

\begin{example}\label{realnumbers}
Let $M=[0,2]\subset\RR$ with the usual topology. This space is meant to represent a two meter long ruler whose markings range from 0 to 2 meters, such that each $x\in M$ is a position measurement (made with an infinitely sharp pointer). In practice no measurement can be infinitely precise, and the best we can do is, when placing the pointer on the marking that corresponds to $x\in M$, to record an open set $U$ that contains $x$. In other words, the open neighborhoods of $x$ are the physical properties that are compatible with $x$. The more precise a measuring device is, the smaller the open sets it will be able to register, but no canonical open set exists for $x$, and certainly no least open set. So having $x\in U$ should be read as ``$U$ can be the physical property recorded by a measuring device that performs the measurement $x$.''
\end{example}

\begin{example}\label{Mspinhalf}
Consider Schr\"odinger's electron, as described in section~\ref{schrelectron}, letting $\mathcal Z$ be the following set of measurements:
\[
\mathcal Z = \{\boldsymbol z^{\spindown},\boldsymbol z^{\spinup},\boldsymbol z\}.
\]
Writing $U^{\spinup}$ and $U^{\spindown}$ for the properties that correspond to the information recorded on the target, $U^{\spinup}$ means that the electron can go up, and $U^{\spindown}$ means that it can go down. By definition, a run of $\boldsymbol z$ can yield either deflection, so $\boldsymbol z$ is compatible with both
$U^{\spinup}$ and $U^{\spindown}$. However, $\boldsymbol z^{\spinup}$ is not compatible with $U^{\spindown}$, and $\boldsymbol z^{\spindown}$ is not compatible with $U^{\spinup}$. So the topology on $\mathcal Z$ consists of the open sets
\[
\topology(\mathcal Z) =\bigl\{
\emptyset,\{\boldsymbol z\},U^{\spindown},
U^{\spinup},\mathcal Z
\bigr\},
\]
where $U^{\spindown}=\{\boldsymbol z^{\spindown},\boldsymbol z\}$ and
$U^{\spinup}=\{\boldsymbol z^{\spinup},\boldsymbol z\}$. Contrary to Example~\ref{realnumbers}, the space of measurements $\mathcal Z$ is not Hausdorff. The property $\{\boldsymbol z\}$, which is the intersection
$U^{\spindown}\cap U^{\spinup}$, can be recorded by running the measurement $\boldsymbol z$ at least twice until two of the runs have yielded different deflections.
\end{example}

\paragraph{Locales.} Any topology $\topology(M)$ is a \emph{locale} (a ``pointless space''~\cite{pointless}), by which is meant a partially ordered set $L$ for which every subset $S\subset L$ has a \emph{join} (\emph{supremum}) $\V S$, and thus also a \emph{meet} (\emph{infimum}) $\bigwedge S$ (\ie, $L$ is a \emph{complete lattice}, or \emph{sup-lattice}), which moreover satisfies the \emph{infinite distributivity} law
\[
U\wedge\V_i V_i = \V_i U\wedge V_i
\]
for all $U\in L$ and all families $(U_i)$ in $L$,
where $\wedge$ represents \emph{binary} meet.
For $L=\topology(M)$ the order is given by inclusion, and the arbitrary joins $\V_i U_i$ and binary meets $U\wedge V$ are, respectively, given by arbitrary unions $\bigcup_i U_i$ and binary intersections $U\cap V$.

For instance, the order structure of the locale $\topology(\mathcal Z)$ of Example~\ref{Mspinhalf} is depicted in the following lattice diagram, where we see that $\{\boldsymbol z\}=U^{\spindown}\wedge U^{\spinup}$ and $\mathcal Z= U^{\spindown}\vee U^{\spinup}$ (and $\vee$ stands for \emph{binary join}):
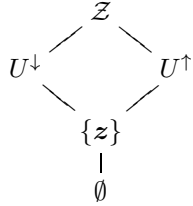
\[
\xymatrix@=8pt{
&\mathcal Z\ar@{-}[dl]\ar@{-}[dr]\\
U^{\spindown}\ar@{-}[dr]&&U^{\spinup}\ar@{-}[dl]\\
&\{\boldsymbol z\}\ar@{-}[d]\\
&\emptyset
}
\]

The topology of a topological space is the prototypical example of a locale, although not all locales are of this form.

\paragraph{Geometric logic.} Locales can be given a logical interpretation, as models of \emph{propositional geometric logic}~\cite{Vi07}: a meet $U\wedge V$ is a logical \emph{conjunction} of $U$ and $V$, and a join $\V_i U_i$ is a logical \emph{disjunction} of the family $(U_i)$. There is an asymmetry because disjunctions can be of arbitrary arity, whereas conjunctions are only finitary.

This can be aptly illustrated in the case of the locale $\topology(M)$ of a space of measurements $M$: the condition that $m\in U_i$ for all physical properties $U_i$ in a given family $(U_i)$ is equivalent to $m\in\bigcap_i U_i$, and in this sense $\bigcap_i U_i$ is the conjunction of the properties $U_i$; however, in general $\bigcap_i U_i$ is not an open set, so not a physical property. On the other hand, the meet $\bigwedge_i U_i$ always exists: it is the interior of $\bigcap_i U_i$. However, we may have $m\in U_i$ for all $i$ but $m\notin\bigwedge_i U_i$, so in general $\bigwedge_i U_i$ is not a good notion of conjunction of the properties $U_i$.

Summing up: in general, the only meets of $\topology(M)$ that should be regarded as conjunctions are the finitary ones because they coincide with intersections. But the condition $m\in\bigcup_i U_i$ is equivalent to the existence of $i$ such that $m\in U_i$, so infinitary disjunctions always exist.

\paragraph{Finite observability.} The asymmetry between the arities of conjunctions and disjunctions justifies the claim that open sets can be regarded as \emph{finitely observable properties}: in the absence of more specific information, the verification that a measurement $m$ is compatible with the conjunction $U\wedge V$ is done by running the measurement $m$ at least twice in order to record both $U$ and $V$ in different runs (\cf\ $U^{\spindown}\cap U^{\spinup}$ in Example~\ref{Mspinhalf}). If the resources used by each run of $m$ (such as time, or energy) are lower bounded by fixed positive quantities, an infinite number of runs is impossible because it would consume infinitely many resources. Hence, the logic of observable properties only contains finitary conjunctions.
However, in order to verify that $m$ belongs to the disjunction $\V_i U_i$ all that is needed is to verify that it belongs to one of the disjuncts $U_i$.

Accordingly, from here on the locale structure of the topology $\topology(M)$ of a space of measurements $M$ will be regarded as its \emph{logic} of physically observable properties. The greatest open set $M$ is the \emph{trivial property}, which conveys no information, and the least open set $\emptyset$ is the \emph{impossible property}, which can never be measured.

\paragraph{Sobriety.} A \emph{homomorphism} of locales $f:L\to L'$ is a mapping between locales that respects the logical interpretation discussed above: for arbitrary families $(U_i)$ in $L$ (including the empty family) and arbitrary $U,V\in L$, the conditions
\[
f\bigl(\V_i U_i)=\V_i f(U_i),\quad f(U\wedge V) = f(U)\wedge f(V),\quad f(1_L)=1_{L'}
\]
hold, 
where $1_L$ and $1_{L'}$ denote the greatest elements of the corresponding locales; and an \emph{isomorphism} is a bijective homomorphism. In particular, a homomorphism $p:L\to\boldsymbol 2$ (the locale $\boldsymbol 2$ equals $\{0,1\}$ with $0<1$) is often referred to as a \emph{point} of $L$ and it can be regarded as a \emph{truth-valuation} that assigns to each ``proposition'' $U\in L$ the truth value $1$ (true) or $0$ (false).

For example, if $m$ is a measurement in a space of measurements $M$, a point $p_m:\topology(M)\to\boldsymbol 2$ is defined by the condition
\[
p_m(U)=1\iff m\in U.
\]
This particular truth-valuation assigns ``true'' exactly to those properties $U$ that are compatible with $m$.

If every point $p:\topology(M)\to\boldsymbol 2$ arises in this manner from a unique measurement $m$, the space $M$ is said to be \emph{sober}. Equivalently, $M$ is sober if and only if every irreducible closed set is the topological closure of a singleton $\overline{\{m\}}$ for a unique measurement $m$. Example: any Hausdorff space.

Requiring $M$ to be a sober space implies that no distinctions between measurements can be made except those warranted by physical properties, so $M$ is a $T_0$ space, and that for any truth-valuation $p:\topology(M)\to\boldsymbol 2$ there is a measurement $m$ that represents it. The latter is a principle of consistency: if from the logical structure of properties it is inferred (albeit possibly transfinitely) that a certain measuring process exists, then such a process should really exist. So from here on all the spaces of measurements in this paper will be assumed to be sober (but not Hausdorff, as will be clear below).

\subsection{Measurement order}

Let $M$ be a sober space of measurements. The \emph{specialization order} of $M$ is given for all $m,n\in M$ by
\[
m\le n \iff m\in\overline{\{n\}},
\]
and it is a partial order which is complete under the formation of joins of directed sets (a set $D$ is directed if it is nonempty and every pair $m,n\in D$ has an upper bound in $D$). For instance, the two meter ruler of Example~\ref{realnumbers} is a Hausdorff space, so its specialization order is discrete.

The condition $m\le n$ holds if and only if $m\in U$ implies $n\in U$ for all properties $U$; that is, $n$ is compatible with at least as many physical properties as $m$. In computer science this would usually be taken to mean that $n$ is a state of a computation which is ``more complete'' than $m$ because more properties have been determined for $n$ than for $m$. However, for measurements the rationale is somewhat different, as the following example illustrates.

\begin{example}\label{specorderspin}
Considering the open sets for spin measurements along $z$ as described in Example~\ref{Mspinhalf}, we obtain $\boldsymbol z^{\spindown}\le \boldsymbol z$ and $\boldsymbol z^{\spinup}\le \boldsymbol z$, so the specialization order of the space $\mathcal Z$ can be represented by the following diagram:
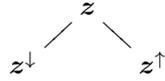
\[
\xymatrix@=10pt{
&\boldsymbol z\ar@{-}[rd]\ar@{-}[ld]\\
\boldsymbol z^{\spindown}&&\boldsymbol z^{\spinup}
}
\]
In an obvious sense $\boldsymbol z$ is \emph{less} determined than either $\boldsymbol z^{\spindown}$ or $\boldsymbol z^{\spinup}$, since the former has more \emph{potential} properties then either of the other two measurements: a run of $\boldsymbol z$ can both yield an upwards deflection and a downwards one.
\end{example}

\subsection{Disjunctions}\label{sec:disj}

In Example~\ref{specorderspin} we see a situation where two measurements, $\boldsymbol z^{\spindown}$ and $\boldsymbol z^{\spinup}$, have a join:
$
\boldsymbol z=\boldsymbol z^{\spindown}\vee\boldsymbol z^{\spinup}
$.
Due to the closed box version of ``Schr\"odinger's electron'' it is tempting to think of the measurement $\boldsymbol z$ as a ``quantum superposition'' of $\boldsymbol z^{\spindown}$ and $\boldsymbol z^{\spinup}$, although this is a superposition only in a logical sense in which amplitudes or probabilities are not involved: a measurement simply belongs or does not belong to a superposition. 

The join $\boldsymbol z^{\spindown}\vee\boldsymbol z^{\spinup}$ can also be regarded as a disjunction: it can be read ``$\boldsymbol z^{\spindown}$ \emph{or} $\boldsymbol z^{\spinup}$,'' since each run of $\boldsymbol z$ is, in terms of the resulting classical information, either a run of $\boldsymbol z^{\spindown}$ or one of $\boldsymbol z^{\spinup}$. This description makes the join of two measurements similar to a  disjunction of propositions in classical logic, where either disjunct may hold but we just do not know which.

Partly due to mathematical convenience, I shall adopt the completeness assumption that for every pair of measurements $m,n\in M$ there is a disjunction of $m$ and $n$, represented by a join $m\vee n$ as above, even if it may not always be clear how certain disjunctions might be physically realized.
Then, since in a sober space $M$ every directed set has a join, the existence of binary joins implies that every nonempty subset has a join.

Moreover, I shall  assume that $M$ contains the join $0 = \V\emptyset$ of the empty set, which is the least element in the specialization order. It follows that $M$ is a complete lattice.
The only property that is compatible with $0$ is $M$, which conveys no information, so $0$ will be called the \emph{impossible measurement}.

Elsewhere~\cite{msnew} it will be shown that the interpretation of binary joins as disjunctions forces the operation $\vee: M\times M\to M$ to be continuous, so from here on spaces of measurements will be assumed to be of the following type:
\begin{definition}
By a \emph{sober lattice} is meant a sober topological space $M$ whose specialization order has a least element $0$ and a join $m\vee n$ for each $m,n\in M$, such that the binary join operation
$\vee:M\times M\to M$ is continuous. (Notice that $M$ is never Hausdorff unless $M=\{0\}$.)
\end{definition}

A consequence of taking spaces of measurements to be sober lattices is that $m\vee n$ is compatible with a property $U$ if and only if there are properties $V$ and $W$, respectively of $m$ and $n$, whose conjunction $V\wedge W$ implies $U$~\cite{msnew}. This makes precise the idea that any property of a disjunction of measurements can be verified via multiple runs that elicit properties of the disjuncts.

Further clarification of the distinction between the ``quantum superposition'' and the ``logical disjunction'' interpretations of binary joins cannot be expressed in terms of the joins alone, but rather it resides in other lattice-theoretical structure and properties, to be addressed in section~\ref{sec:classical} (\cf\ Example~\ref{exm:idlemeasurement}).

\section{Measurement algebra}

\subsection{Product}\label{sec:comp}

If in addition to a disjunction we were to postulate the existence of conjunctions of measurements, should we interpret ``$m$ \emph{and} $n$'' as a measurement that consists of performing both $m$ and $n$? If so, in which order? For instance, in a physical experiment we may want to measure the spin of an electron along $z$ and \emph{after that} measure the spin along $x$ of \emph{the same electron}, which can be done by setting up the apparatus so that the electron traverses two magnetic fields one after the other. Such a succession of measurements can be regarded as a conjunction, but one which is extended in time and thus is not necessarily idempotent or commutative.

For a more pedestrian example, think of the experiment that consists of pressing a key of a computer keyboard while watching the corresponding character appear on the computer screen. Pressing the key ``a'' twice produces the string ``aa'', and pressing ``a'' and then ``b'' produces the string ``ab,'' which is not what we obtain if we press ``b'' and then ``a.'' So we see that there are ``conjunctions'' ``aa,'' ``ab,'' and ``ba,'' but ``a''$\neq$ ``aa'' and ``ab''$\neq$``ba'' (cf.\ \cites{M86,Re01,AV93}).

When a measurement $m$ is followed in the manner just suggested by another measurement $n$, I shall denote the composed measurement by $nm$ (notice the reversed order), and call it the \emph{product} of $n$ and $m$. It will be assumed that all pairs of measurements can be composed, but we can express that a composition $nm$ is meaningless by writing $nm=0$.

Let us look at the properties of the product.

\begin{description}
\item[Associativity:] I shall make no distinction between $(mm')m''$ and $m(m'm'')$.
\item[Absorption:] Naturally any composition that involves the impossible measurement must be impossible itself, \ie, we must have $m0=0m=0$.
\item[Continuity:] Analogously to disjunctions, the product will be assumed to be continuous.
\item[Distributivity:] If $m$, $n$ and $n'$ are measurements, the product $m(n\vee n')$ will be identified with $mn\vee mn'$, and $(n\vee n')m$ will be identified with $nm\vee n'm$.
\end{description}
The justification for distributivity is based on the interpretation of joins as disjunctions: an execution of $m$ followed by \underline{either $n$ or $n'$} is the same as one of either \underline{$m$ and then $n$} or \underline{$m$ and then $n'$}. But, even for a join $n\vee n'$ thought of as a quantum superposition, we can regard distributivity as a weak form of bilinearity. Section~\ref{Cstar} will provide examples.

Distributivity means that the product preserves binary joins in each variable, and continuity implies that it preserves joins of directed sets in each variable. Adding absorption, which means that the empty join is likewise preserved, we conclude that the product preserves arbitrary joins in each variable, so spaces of measurements are \emph{quantales}~\cite{Rosenthal1}.

\subsection{Involution}

The last ingredient of the definition of space of measurements is an \emph{involution} which to each measurement $m$ assigns its \emph{involute} $m^\inv$, satisfying the following properties for all measurements $m$ and $n$:
\begin{itemize}
\item $(-)^\inv$ is continuous.
\item $(m^\inv)^\inv=m$.
\item $(mn)^\inv=n^\inv m^\inv$.
\end{itemize}
The first two conditions imply that the involution is a homeomorphism, so it is also an order isomorphism, and thus the space of measurements is an \emph{involutive} quantale.

The involution can be regarded as a formal time reversal such that $m^\inv$ is  $m$ backward in time. This does not mean that measurements are meant to be reversible processes (they are not ``unitary transformations'') but only that, analogously to any operator in a C*-algebra, they have ``adjoints'' (\cf\ section~\ref{Cstar}).

While in practice it may not always be clear what $m^\inv$ represents physically for a given measurement $m$, the involution has useful mathematical consequences. Moreover, it provides a way of describing reversibility in situations where that can have a meaning: for a measurement $m$ to be considered reversible, in the sense that it can be ``undone'' by $m^\inv$, we should at least require $mm^*m=m$. Let us call such a measurement \emph{regular}.

A \emph{lax} form of regularity is given by $mm^\inv m\le m$, which implies that $mm^\inv m$ is a special way of performing $m$. But, thinking in terms of reversibility, this would mean that $m^\inv$ is not able to undo $m$ perfectly, as if some trace of the undoing were memorized while performing $mm^\inv m$. Such ``fake'' reversibility can be ruled out by imposing that whenever $mm^\inv m\le m$ the measurement $m$ must be regular. Spaces of measurements that satisfy this condition are stably Gelfand quantales~\cite{GSQS}, so they contain a rich supply of embedded pseudogroups~\cite{SGQ} and Morita equivalence bimodules between them~\cite{msnew}, whose elements can indeed be regarded as reversible transformations.

Piecing together the above ingredients we arrive at our main definition:

\begin{definition}\label{def:ms}
By a \emph{measurement space} is meant a sober lattice $M$ equipped with two continuous operations
\begin{eqnarray*}
(n,m)\mapsto nm&:&M\times M\to M\quad\text{(\emph{composition})},\\
m\mapsto m^*&:& M\to M\quad\text{(\emph{involution})},
\end{eqnarray*}
such that for all $m,n,p\in M$ the following axioms are satisfied:
\begin{enumerate}
\item $(nm)p=n(mp)$\quad (\emph{associativity}),
\item $(n\vee m)p = np\vee mp$\quad (\emph{distributivity}),
\item $0p=0$\quad (\emph{absorption}),
\item $m^{**}=m$\quad (\emph{involution I}),
\item $(nm)^*=m^*n^*$\quad (\emph{involution II}),
\item $mm^*m\le m\Longrightarrow mm^*m=m$\quad (\emph{reversibility}).
\end{enumerate}
(Note that absorption and distributivity exist in both variables due to the involution.)

A \emph{homomorphism} of measurement spaces $h:M\to N$ is a continuous map which for all $m,n\in M$ satisfies
\[
h(0)=0,\ \ h(m\vee n) =h(m)\vee h(n),\ \ h(mn)=h(m)h(n),\ \ h(m^*)=h(m)^*,
\]
and an \emph{isomorphism} is a homomorphism which is also a homeomorphism.
\end{definition}

\subsection{Algebraic quantum theory}\label{Cstar}

Let $A$ be a C*-algebra, and let $\Max A$ be Mulvey's involutive quantale of closed linear subspaces of $A$~\cite{Curacao,MP1}, whose involution is computed pointwise from the involution of $A$ and whose product is defined to be the closure of the linear span of the pointwise product:
\[
PQ = \overline{\spanmap\{ab\st a\in P,\ b\in Q\}}.
\]
This is a stably Gelfand quantale~\cite{QFB}, and it is a sober space if we equip it with the lower Vietoris topology~\cite{RS2}, whose open sets are the unions of finite intersections of the form
$\ps U_1\cap\ldots\cap \ps U_k$,
where for each open set $U$ of $A$ we have
\[
\ps U = \{P\in\Max A\st P\cap U\neq\emptyset\}.
\]
It can be proved that the algebraic operations of $\Max A$ are continuous~\cite{msnew}, so $\Max A$ is a measurement space.

Since in algebraic quantum theory systems are described by operator algebras that contain the observables (or local observables) of the systems, the $\Max A$ construction allows us to associate to any such system a measurement space in a canonical way.

\begin{example}
Let us illustrate this with the Schr\"odinger electron example. Taking the C*-algebra to be $M_2(\CC)$, the algebra of $2\times 2$ complex matrices, the associated measurement space $M$ is $\Max M_2(\CC)$, the space of all the linear subspaces of $M_2(\CC)$, with the lower Vietoris topology. Then the spin measurements along $x$ and $z$ can be described via Pauli spin matrices and projections in an unsurprising way: $\boldsymbol x$ and $\boldsymbol z$ are unital abelian subalgebras generated by the corresponding spin observables, and $\boldsymbol z^{\spindown}$, $\boldsymbol z^{\spinup}$, $\boldsymbol x^{\spindown}$, and $\boldsymbol x^{\spinup}$ are spanned by the matrices that project onto the respective eigenspaces:
\begin{align*}
&\boldsymbol z^{\spinup} = \left\langle\left(\begin{array}{cc}1&0\\0&0\end{array}\right)\right\rangle
\quad\quad \boldsymbol z^{\spindown} =  \left\langle\left(\begin{array}{cc}0&0\\0&1\end{array}\right)\right\rangle\\
\boldsymbol z = D_2(\CC) &=\left\langle \left(\begin{array}{cc}1&0\\0&1\end{array}\right)\ ,\ \left(\begin{array}{cc}1&0\\0&-1\end{array}\right)\right\rangle\\&\boldsymbol x^{\spinup} = \left\langle\left(\begin{array}{cc}1&1\\1&1\end{array}\right)\right\rangle\quad\quad
\boldsymbol x^{\spindown} =  \left\langle\left(\begin{array}{cc}1&-1\\-1&1\end{array}\right)\right\rangle\\
\boldsymbol x &=\left\langle \left(\begin{array}{cc}1&0\\0&1\end{array}\right)\ ,\ \left(\begin{array}{cc}0&1\\1&0\end{array}\right)\right\rangle
\end{align*}
A fragment of the measurement order is this:
\[
\xymatrix@=10pt{
&\boldsymbol x\ar@{-}[dl]\ar@{-}[dr]&&&&\boldsymbol z\ar@{-}[dl]\ar@{-}[dr]\\
\boldsymbol x^{\spindown}\ar@{-}[drrr]&&\boldsymbol x^{\spinup}\ar@{-}[dr]&&\boldsymbol z^{\spindown}\ar@{-}[dl]&&\boldsymbol z^{\spinup}\ar@{-}[dlll]\\
&&& 0
}
\]
Note that $\boldsymbol x\neq\boldsymbol z$, as intended (\cf\ section~\ref{schrelectron}).
\end{example}

\begin{example}\label{exm:localsys}
In algebraic quantum field theory relativity is taken into account by considering C*-algebras $A$ equipped with local C*-systems~\cite{Araki93}*{Ch.\ 4}. Analogously, define a \emph{local measurement system} on a measurement space $M$ to consist of:
\begin{enumerate}
\item An action of the Poincar\'e group by automorphisms on $M$.
\item An involutive subquantale $M(D)$ (a subspace closed under all joins, composition and involution), whose elements are called the \emph{local measurements at $D$}, for each open causally convex bounded region $D$ of Minkowski spacetime, such that:
\begin{enumerate}
\item The family $\{M(D)\}$ is a monotone net in $M$.
\item For all pairs of spatially separated regions $D$ and $D'$ and all local measurements $m\in M(D)$ and $m'\in M(D')$ we have $mm'=m'm$.
\item The Poincar\'e action on $M$ is covariant with respect to the action on the set of regions $D$.
\item Every measurement $m\in M$ is a join of local measurements.
\end{enumerate}
\end{enumerate}
Given any local C*-system $\{A(D)\}$ on a C*-algebra $A$ we obtain a local measurement system $\{\Max A(D)\}$ on $\Max A$ whose Poincar\'e group action is the extension to $\Max A$ of the action on $A$.
\end{example}

This example provides a mathematically straightforward way of bringing relativity to bear on measurement spaces, however in a background dependent form that from a broader point of view may be regarded as a limitation~\cite{Smolin06}.

\subsection{Time and causality}\label{sec:time}

In section~\ref{sec:comp} I introduced the product of measurements as a way of describing ``conjunctions'' that are extended in time, for instance when two Stern--Gerlach apparatuses are placed one after the other in order to be traversed sequentially by the same particle. That simple picture sufficed for introducing the main ingredients of the theory, but it is worth being more careful in order to understand how time and causality are present in the framework, in particular hinting at how relativistic aspects can be accounted for in a background independent way.

For instance, $mn$ may express simply that $n$ was measured before $m$, meaning that a clock indicated that $m$ started after $n$ finished. Then a commutation relation
$
mn=nm
$
can be taken to mean that different clocks yield opposite time order information, and thus that $m$ and $n$ correspond to spatially separated events in Minkowski spacetime. So a similar approach to that of algebraic quantum field theory can be adopted, in particular letting $m$ and $n$ be local measurements in spatially separated regions, as in Example~\ref{exm:localsys}. However, this approach is somewhat rigid, as several labeled copies of otherwise identical measurements need to be introduced in order for some of them to commute. For instance, consider the spin measurements $\boldsymbol x$ and $\boldsymbol z$ labeled by regions: we will have $\boldsymbol x_D\boldsymbol z_{D'}=\boldsymbol z_{D'}\boldsymbol x_{D}$ for two spatially separated regions $D$ and $D'$, but $\boldsymbol x_D\boldsymbol z_{D}\neq \boldsymbol z_{D}\boldsymbol x_{D}$ for consecutive spin measurements on the same electron.

Alternatively, a more flexible and more background independent approach is to interpret $mn$ as a measurement composed of two parts $m$ and $n$ under the knowledge that $n$ is a causal precondition for $m$, such as in the aforementioned composition of Stern--Gerlach apparatuses. The product $mn$ can be represented graphically by the following diagram whose arrow indicates ``is a causal precondition for,''
\[
\xymatrix{m\\n\ar[u]}
\]
and we can also group together larger finite sets of measurements, for instance as in this example where we have two copies of $m$ and two copies of $n$ with the three depicted causal links:
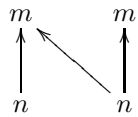
\[
\xymatrix{m&m\\
n\ar[u]&n\ar[lu]\ar[u]}
\]
This N shape defines an operation in four variables which we may write as $N(x,y,z,w)$, here instantiated with $x=y=m$ and $z=w=n$.
Moreover, if the causality in $mn$ is interpreted as stating that $m$ is in the future light cone of $n$, a diagram like the one above can be regarded as a finite causal set~\cite{BLMS87} (a finite strictly partially ordered set) together with a labeling of its points (events) with measurements. Any finite causal set with $k$ elements can be thus regarded as a $k$-ary operation on measurements.

For instance, consider a measurement that contains two parts $m$ and $n$ which are observed to be independent (for instance by using two clocks). This is represented by the labeled causal set
\[
\xymatrix{m&n},
\]
which I will denote by $m\| n$. The operation $\|$ should be associative and commutative and relate to the topology and the order in the same way as the product. It allows us to express independence of $m$ and $n$ without resorting to the commutation relation $mn=nm$; for instance we can write $\boldsymbol x\|\boldsymbol z$ for two independent spin measurements without needing to introduce several labeled copies of $\boldsymbol x$ and $\boldsymbol z$.

The new operations must also relate to the involution in a way that respects the law $m^{**}=m$; \eg, the obvious generalization of $(mn)^*=n^* m^*$ consists of replacing each label $m$ by $m^*$ while inverting the arrows, as in this example:
\[
\left(\vcenter{\xymatrix{m&m\\
n\ar[u]&n\ar[lu]\ar[u]}}\right)^* = \vcenter{\xymatrix{m^*&m^*\\
n^*\ar@{<-}[u]&n^*\ar@{<-}[lu]\ar@{<-}[u]}}
\]

These ideas suggest that information about independence and causality in a space of measurements can be encoded by equipping a sober lattice $M$ with continuous $k$-ary algebraic operations $f:M^k\to M$, indexed by finite causal sets, which preserve joins in each variable and relate to the involution as just described. Moreover, the fact that the specialization order can often be regarded as a ``containment relation'' (\cf\ section~\ref{sec:classical} below) suggests connections to the causal sites of~\cite{CC05}.

It is natural to expect that (some) theories of quantum gravity should be the place where to look for examples of measurement spaces with such additional operations, but this will not be considered here. In the remainder of the paper, by a measurement space will always be meant a space according to Definition~\ref{def:ms}.

\section{Classical measurements}\label{sec:classical}

\subsection{Conjunctions}\label{sec:conjunctions}

Since measurement spaces are complete lattices, arbitrary meets exist for free.
A meet $m\wedge n$ has fewer compatible physical properties than either $m$ or $n$, so it is more determined, but it is the least determined measurement that is both more determined than $m$ and $n$. We may read it as ``$m$ \emph{and} $n$,'' the measurement that is exactly \emph{both} $m$ and $n$ --- in other words, $\wedge$ is interpreted as \emph{logical conjunction} of measurements. For instance, in the electron spin example the meet of $\boldsymbol z^{\spindown}$ and $\boldsymbol z^{\spinup}$ is the impossible measurement $0$, in accordance with the fact that no spin measurement of a single particle can yield both an upwards and a downwards deflection simultaneously.

In any lattice the binary meets relate to binary joins via the following lax distributivity law:
\[
(p\wedge m)\vee (p\wedge n)\le p\wedge(m\vee n).
\]
Hence, regarding a meet of measurements $p\wedge q$ as a selection of the ``$p$-component'' of $q$ (as one would do when thinking of projections on a Hilbert space), if the above inequality is strict it follows that the $p$-component of the disjunction $m\vee n$ is less determined than the disjunction of the $p$-components of $m$ and $n$. In other words, it may be impossible to verify an arbitrary property $U$ of $p\wedge(m\vee n)$ by obtaining in separate runs two properties $V$ and $W$ such that $V\cap W\subset U$, respectively of $p\wedge m$ and $p\wedge n$. Similarly to traditional quantum logics based on orthomodular lattices, such a lack of distributivity is a sign of nonclassical behavior.

Hence, in analogy with the algebraic models of propositional geometric and intuitionistic logic, one of the ingredients that will be required for considering a space of measurements to be classical is \emph{distributivity}:
\[
(p\wedge m)\vee (p\wedge n)= p\wedge(m\vee n).
\]

\begin{example}\label{exm:idlemeasurement}
Consider the matrix representations of the spin measurements $\boldsymbol x$ and $\boldsymbol z$ given in section~\ref{Cstar}. We have the following instance of non-distributivity:
\[
\boldsymbol x\wedge (\boldsymbol z^\spindown\vee \boldsymbol z^\spinup)=\boldsymbol x\wedge \boldsymbol z = e\neq 0=(\boldsymbol x\wedge \boldsymbol z^\spindown)\vee(\boldsymbol x\wedge \boldsymbol z^\spinup),
\]
where $e$ denotes the span $\langle I\rangle$ of the identity matrix $I$.
It is worth looking at an intuitive interpretation of this. For instance, $\boldsymbol x\wedge \boldsymbol z^\spindown$ would be a measurement of spin along $x$ during which a downwards vertical deflection is observed. This is impossible, in accordance with the fact that in $\Max M_2(\CC)$ we have $\boldsymbol x\wedge \boldsymbol z^\spindown=0$. Similarly, we have $\boldsymbol x\wedge \boldsymbol z^\spinup=0$. However, $\boldsymbol x\wedge \boldsymbol z\neq 0$. In particular, this measurement contains the identity matrix:
\[
I\in \boldsymbol x\wedge \boldsymbol z.
\]
Hence, any observable property that contains $e$ also contains $\boldsymbol x$ and $\boldsymbol z$.

Since $e$ is the multiplicative unit of $\Max M_2(\CC)$, it can be interpreted as a trivial measurement that does nothing to the system being observed (not to be confused with the top measurement $1$, which is compatible with every possible observable property). With this point of view in mind, the inequalities
\[
e\le \boldsymbol x\quad\text{and}\quad e\le \boldsymbol z
\]
mean that the act of doing nothing is a particular way of performing both $\boldsymbol x$ and $\boldsymbol z$; in other words, any observable property which may be obtained by just sitting idly is necessarily compatible with both $\boldsymbol x$ and $\boldsymbol z$. Note, however, that $e\wedge \boldsymbol z^{\spinup}=0$, $e\wedge \boldsymbol z^{\spindown}=0$, $e\wedge \boldsymbol x^{\spinup}=0$ and $e\wedge \boldsymbol x^{\spindown}=0$, expressing that a measurement of an actual deflection is incompatible with doing nothing.

So we see that, despite the fact that every property of $\boldsymbol z$ can be measured by intersecting properties obtained in multiple runs of $\boldsymbol z^{\spindown}$ and $\boldsymbol z^{\spinup}$ (\cf\ section~\ref{sec:disj}), some properties of $\boldsymbol z$ can also be known by other means, some even just ``tautologically.'' This is consistent with the logical interpretation~\cite{MP1} of the quantale $\Max A$ of a unital C*-algebra $A$, where $e$ is interpreted as the ``true'' proposition, and an inequality such as $e\le m$ means that $m$ ``is'' true. So, in this sense, both $\boldsymbol x$ and $\boldsymbol z$ are true in $\Max M_2(\CC)$, but more specific measurements like $\boldsymbol z^{\spinup}$ are not.

This also suggests good examples of joins that represent ``quantum superpositions'' (\cf\ section~\ref{sec:disj}): if for a given measurement space $M$ with unit $e\neq 0$ we have $e\le m\vee n$ then $m\vee n$ can be regarded as a superposition in the style of Schr\"odinger's cat if we regard $e$ (doing nothing) as simply keeping the box closed, thereby obtaining no classical information about the status of the cat. However, this interpretation of $m\vee n$ is ruled out if $M$ is a distributive lattice and if, similarly to the electron example, we have $e\wedge m=e\wedge n=0$, for then we must also have $e\wedge(m\vee n)=0$.
\end{example}

\subsection{Continuous lattices and the Scott topology}\label{sec:continuouslattices}

Let us adopt the stance that classical measurements do not conflict with the common intuition according to which physical systems have an intrinsic and observer-independent existence. Then obtaining an observable property $U$ from a measurement $m$ can simply be regarded as \emph{learning more about the observed system in a way that never conflicts with previously obtained information}: each measurement can be regarded as data about the observed system, with $m\le n$ meaning that $n$ is ``more data'' because more properties are known about it.

Moreover, each measurement can be used as an input for computing other data via some finite computational process, as one usually does in experimental setups by means of data acquisition and processing systems. Such a process itself can be considered to be part of a measurement. But while some measurements (such as some infinite joins) may carry infinite information, computations based on them need to proceed on the basis of finite approximations (such as when computing a real-valued function on the real numbers). Following this argument carefully it is proved~\cite{Stoy}*{chapter 6} that the spaces on which such computational processes act must be \emph{countably-based continuous lattices}~\cite{bigdomainbook}.

From here on it will be assumed that the specialization order of a space of classical measurements is such a lattice. The topology of physical properties in this case should be the \emph{Scott topology}, which for a continuous lattice is the finest sober topology, and moreover allows the aforementioned computational processes to be represented by continuous functions --- the open sets are the upper-closed sets $U\subset M$ that are inaccessible by joins of directed sets:
\begin{itemize}
\item $m\in U$ and $m\le n$ implies $n\in U$;
\item $\V D\in U$ for a directed set $D$ implies $D\cap U\neq\emptyset$.
\end{itemize}

\subsection{Classical measurement spaces}

It is well known~\cite{bigdomainbook} that distributive continuous lattices are necessarily \emph{locally compact locales}; that is, each such lattice is isomorphic to the locale of open sets $\topology(X)$ of a (unique up to homeomorphism) sober topological space $X$ which is locally compact in the strong sense that for all points $x\in X$ and open sets $U$ such that $x\in U$ there is another open $V$, and a compact set $K$, such that $x\in V\subset K\subset U$.

Therefore, we see that by putting together the two requirements for classicality described respectively in section~\ref{sec:conjunctions} (a logical condition) and section~\ref{sec:continuouslattices} (a computational condition), we are led to the conclusion that classical measurements ought to look like they are open sets of underlying topological spaces, moreover whose topologies are ``reasonable.'' This is interesting because it resonates with the intuition that classical systems have underlying spaces of ``real'' states, and leads to the following definition:

\begin{definition}
A measurement space $M$ is \emph{classical} if the following three conditions hold:
\begin{enumerate}
\item The specialization order of $M$ defines a countably-based locally compact locale; equivalently, it is isomorphic to the topology $\topology(X)$ of a second-countable sober locally compact space $X$.
\item The topology of physical properties $\topology(M)$ is the Scott topology.
\item $m\le mm^*m$ for all $m\in M$.
\end{enumerate}
Moreover, $M$ is \emph{local} if $mn=m\wedge n$ and $m^*=m$ for all $m,n\in M$.
\end{definition}

Note that the continuity of joins, compositions and the involution is ensured by well known properties of the Scott topology: functions are continuous if and only if they preserve joins of directed sets; functions in several variables are continuous if and only if they are continuous in each variable separately.

The third condition of the definition is a strong version of the reversibility axiom of Definition~\ref{def:ms}. It expresses the idea, valid for state-based systems, that any product $m^*m$, which is an execution of $m$ followed by its time reversal, should at least be able to carry the system back to its original state (but possibly other states, too), and thus $mm^*m$ can induce at least as many state transitions as $m$.

\begin{example}\label{exm:labwall}
Let $\wall$ be a rectangle $[a,b]\times[c,d]\subset \RR^2$ ($a<b$ and $c<d$) that represents the surface of a wall of the lab where we are making measurements.
The wall is observed by visual inspection, and each photon that hits our retinas carries information about a region, no matter how small, which can be modeled by an open set $U\subset \wall$. No visual inspection will have high enough resolution for actual points to be seen (which is good because this sidesteps the question of whether points have a physical meaning at all), and the best we can do is to perform successively sharper measurements $U_1$, then $U_2$, then $U_3$, etc., so that at each stage we will have performed the product measurement $U_k\cdots U_1=U_k\cap\ldots\cap U_1$.

Therefore such measurements are described by the local measurement space $\topology(\wall)$. The points of $\wall$ play the role of states, but these are only derived constructions because the actual measurements are open sets.

This example is entirely compatible with the usual description of classical systems by abelian C*-algebras: by the Gelfand--Naimark representation theorem, $\topology(\wall)$ is isomorphic to the locale $I(C(\wall))$ of closed ideals of $C(\wall)$; and the topology of $I(C(\wall))$ as a subspace of $\Max C(\wall)$ coincides with the Scott topology~\cite{msnew}, so $\topology(\wall)$ and $I(C(\wall)$ are isomorphic measurement spaces.
\end{example}

\subsection{Groupoids}

An important example of classical measurement space is the quantale $\opens(G)$ of~\cite{Re07}, equipped with the Scott topology, in those cases where $G$ is a sober second-countable locally compact groupoid with open domain map (\eg, a Lie groupoid). Concretely, the measurements are the open sets of the space of arrows $G_1$, the product is pointwise composition of arrows, and the involution is pointwise inverse. This coincides with the local measurement space $\topology(G_0)$ if $G$ is just a space $G_1=G_0$.

\begin{example}
Geometric structure of the lab wall of Example~\ref{exm:labwall} can be described by letting $\wall$ be the object space $G_0$ of a Lie groupoid $G$ that encodes local symmetries of the wall, \eg, such that each arrow $g:x\to y$ is a specification of how to map neighborhoods of $x$ into ``similar'' neighborhoods of $y$ (\cf\ \'etale groupoids associated to inverse semigroups, in~\cites{Armstrongetal21} and references therein). The measurement space $\opens(G)$ is classical, and again $\wall$ plays the role of state space, now equipped with ``dynamics'' provided by the groupoid structure.
\end{example}

\section{Observers}

\subsection{Approximated measurements}

Let $M$ be a measurement space, meant to describe everything that can possibly be measured about a given physical system. Any entity, or experimental apparatus, which ``observes'' the system should be able to perform some of the measurements in $M$, but usually not all of them. So a basic ingredient of any mathematical definition of observer should be a subset $\opens\subset M$ which is meant to contain the whole repertoire of measurements that can be performed by a given measurer. Presumably, observers should be capable of applying to their own measurements the entire set of algebraic operations of measurement spaces, such as composition, so $\opens$ will be assumed to be an involutive subquantale of $M$ (\cf\ subquantales $M(D)$ of Example~\ref{exm:localsys}). Being closed under joins guarantees that $\opens$ is a sober space in the relative topology, so it is a measurement space in its own right.

Now let $m\in M\setminus\opens$. Although $m$ is not a measurement that the observer can perform, it is nevertheless a physical event of the system; in particular, $m$ can be a measurement performed by another observer. Let $\opens_A$ and $\opens_B$ be two involutive subquantales of $M$, corresponding to observers Alice and Bob, respectively. Even though Bob has no direct access to all of Alice's measurements, there can be communication between them through a physical channel, possibly involving some amount of information processing. The latter can be, say, a direct electrical connection between Alice's and Bob's measuring devices, or a lengthy procedure that requires Bob to read Alice's scientific papers, etc. But the important thing to retain is that Bob may have indirect access to Alice's experiments and may be able to represent them in terms of his own repertoire of measurements, even if only in an approximated fashion, and this fact ought to be taken into account in any mathematical definition of observer. Of course, there can be different ways of approximating Alice's measurements in terms of Bob's repertoire, which means that the subquantale $\opens_B$ alone is not a good mathematical definition of Bob-the-observer.

Let us assume that each such approximation is described by an  \emph{approximation function} $\af:\opens_A\to\opens_B$ that satisfies (at least) the following properties.
\begin{enumerate}
\item If $m\in\opens_A\cap\opens_B$ then $\af(m)=m$; that is, if $m$ is one of Bob's measurements then it is interpreted as $m$ itself.
\item Each disjunction of measurements is approximated by the disjunction of their approximations: $\af(m\vee n)=\af(m)\vee \af(n)$ for all $m,n\in\opens_A$.
\item The reversal of a measurement is approximated by the reversal of its approximation: $\af(m^*)=\af(m)^*$ for all $m\in\opens_A$.
\item The function $\af$ is continuous. This expresses the idea that the translation $\af$ is ``finite'' because $\af^{-1}(U)$, which is a finitely observable property, is measured by performing the translation specified by $\af$ and then observing the property $U$.
\end{enumerate}
An immediate consequence of these conditions is that $\af$ preserves arbitrary joins (it preserves $0$ because the latter is contained in all the subquantales of $M$). However, in general $\af$ will not be required to preserve composition of measurements, so it is not a homomorphism of quantales. The following example shows that such a requirement would be too strong.

\begin{example}\label{exm:OxOz}
Consider the algebra $M_2(\CC)$ of spin measurements on a spin-$\frac 1 2$ particle, as in section~\ref{Cstar}. The locale
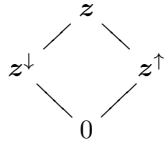
\[
\xymatrix@=10pt{
&\boldsymbol z\ar@{-}[dl]\ar@{-}[dr]\\
\boldsymbol z^{\downarrow}\ar@{-}[dr]&&\boldsymbol z^{\uparrow}\ar@{-}[dl]\\
&0
}
\]
is a local measurement space $\opens_z$ embedded in $M=\Max M_2(\CC)$.
Similarly, let $\opens_x$ be the local measurement space in $M$ that corresponds to measurements of spin along $x$.
Suppose Bob makes the measurements in $\opens_z$ and Alice makes those of $\opens_x$. Whenever Alice communicates to Bob that she has measured $\boldsymbol x^{\uparrow}$, and Bob subsequently (on the same electron) performs a measurement of spin along $z$, he will sometimes obtain $\boldsymbol z^{\uparrow}$ and other times $\boldsymbol z^{\downarrow}$. So he concludes that the best approximation of $\boldsymbol x^{\uparrow}$ is the disjunction $\boldsymbol z=\boldsymbol z^{\uparrow}\vee\boldsymbol z^{\downarrow}$. The same reasoning applies if Alice measures $\boldsymbol x^{\downarrow}$, so the approximation map $\af:\opens_x\to\opens_z$ is as follows:
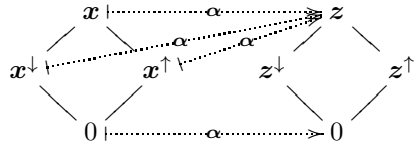
\[
\xymatrix@=10pt{
&\boldsymbol x\ar@{-}[dl]\ar@{-}[dr]\ar@{|.>}[rrrr]|{\af}&&&&\boldsymbol z\ar@{-}[dl]\ar@{-}[dr]\\
\boldsymbol x^{\spindown}\ar@{-}[dr]\ar@{|.>}[urrrrr]|{\af}&&\boldsymbol x^{\spinup}\ar@{-}[dl]\ar@{|.>}[urrr]|{\af}&&\boldsymbol z^{\spindown}\ar@{-}[dr]&&\boldsymbol z^{\spinup}\ar@{-}[dl]\\
&0\ar@{|.>}[rrrr]|{\af}&&&&0
}
\]
Here we see that compositions of measurements are not in general preserved by $\af$, for we have
$\af(\boldsymbol x^{\uparrow}\boldsymbol x^{\downarrow})=\af(0)=0$
but
$
\af(\boldsymbol x^{\uparrow})\af(\boldsymbol x^{\downarrow})=\boldsymbol z\boldsymbol z=\boldsymbol z\neq 0$.
\end{example}

\subsection{Definition of observer}

A special instance of the above ideas arises when an observer with subquantale $\opens$ is a restriction of another observer with subquantale $\opens'$, in the sense that $\opens$ is itself an involutive subquantale of $\opens'$. Let us simply assume that $\opens'=M$ from here on. Then, from the conditions imposed on approximation functions, it immediately follows that any such function
$
\af:M\to\opens
$
must be a topological retraction, albeit not, as already mentioned, a product preserving retraction.

But in this situation there is additional algebraic structure that can be taken into account, namely both $M$ and $\opens$ have defined on them operations of left and right multiplication by elements of $\opens$ (this makes them \emph{$\opens$-$\opens$-bimodules}). Let $\omega\in\opens$ and $m\in M$. The composition $m\omega$ is read ``$\omega$ and then $m$'' by the observer that corresponds to the whole of $M$. Since $\omega\in\opens$, we may regard $\omega$ as a ``state preparation'' done by the observer associated to $\opens$; that is, performing $\omega$ is meaningful from the point of view of $\opens$, and thus $m\omega$ can be interpreted by the observer associated to $\opens$ as ``$\omega$ and then something that approximates $m$.'' In other words, we ought to have
$\af(m\omega)=\af(m)\omega$.
(Hence, $\af$ is a homomorphism of right $\opens$-modules, and, since $\af$ preserves the involution, it is a homomorphism of $\opens$-$\opens$-bimodules.) These ideas lead to the following definition:

\begin{definition}
Let $M$ be a measurement space. By an \emph{observer context} of $M$ will be meant a pair $(\opens,r)$ formed by a subset $\opens\subset M$ closed under arbitrary joins, together with a topological retraction $r:M\to\opens$ that satisfies the following conditions for all $m,n\in M$ and $\omega\in\opens$:
\begin{enumerate}
\item $r(m\vee n)=r(m)\vee r(n)$, \label{arbjoins}
\item $r(m^\inv)=r(m)^\inv$,
\item $r(m\omega)=r(m)\omega$.
\end{enumerate}
(This implies that $\opens$ is an involutive subquantale of $M$, and thus a measurement space.)
We also say that the observer context $(\opens,r)$ is \emph{classical} if $\opens$ is a classical measurement space, and \emph{local} if $\opens$ is a local measurement space.
\end{definition}

\begin{example}\label{exm:OxOz2}
Consider again the involutive subquantale $\opens_z$ of Example~\ref{exm:OxOz}.
A local observer context $(\opens_z,r_z)$ can be obtained by considering the linear surjection
\[
\Theta_z: M_2(\CC)\to\boldsymbol z
\]
which is defined by restricting matrices to the main diagonal:
\[
\Theta_z(A)=\left(\begin{array}{cc} a_{11}&0\\0&a_{22}\end{array}\right).
\]
Then $r_z:M\to \opens$ is
defined by $r_z(P) = \Theta_z(P)\boldsymbol z$ for all $P\in M$.
The approximation map $\af:\opens_x\to\opens_z$ of Example~\ref{exm:OxOz} is the restriction of $r_z$ to $\opens_x$.

More generally, consider any C*-algebra $A$ with a conditional expectation $\Theta:A\to B$ onto an abelian sub-C*-algebra $B\subset A$. A local observer context $(\opens,r)$ of $\Max A$ is obtained by taking $\opens$ to be the locale of closed ideals of $B$ and $r(P)=\overline{\Theta(P)}B$ for all $P\in\Max A$. 
\end{example}

\begin{example}\label{exm:groupoidCstar}
To each matrix $A\in M_n(\CC)$ assign a matrix $S$ of 0s and 1s, called the \emph{support} of $A$, such that $s_{ij}=1$ if and only if $a_{ij}\neq 0$. This defines a map $\supp:M_n(\CC)\to M_n(\boldsymbol 2)$, where the ``algebra'' $M_n(\boldsymbol 2)$ is an involutive quantale whose order is the pointwise order induced by that of the locale $\boldsymbol 2$, and whose multiplication and involution are defined by
\[
(ST)_{ij} = \V_k s_{ik}t_{kj}\quad\quad\text{and}\quad\quad (S^*)_{ij}=s_{ji}.
\]
This quantale is isomorphic to the quantale of binary relations
\[
\mathcal P(\{1,\ldots,n\}\times\{1,\ldots,n\}),
\]
which is the quantale of the pair groupoid $G=\pair(\{1,\ldots,n\})$, and $M_n(\CC)$ is isomorphic to the reduced C*-algebra of $G$. Then for each $U\in\opens(G)$, identified with a matrix $(u_{ij})$ of 0s and 1s, define $\iota(U)$ to be the subspace of matrices $A\in M_n(\CC)$ such that $a_{ij}=0$ if $u_{ij}=0$. This yields an injective homomorphism of involutive quantales $\iota:\opens(G)\to\Max A$, and there is a mapping $r:\Max A\to\opens(G)$ which is the extension of $\supp$ to arbitrary joins. Then the pair $\bigl(\iota(\opens(G)),\iota\circ r\bigr)$ is a classical observer context of $\Max A$.

This is an instance of a more general construction that associates to any second-countable locally compact Hausdorff \'etale groupoid $G$ a canonical observer context $(\opens,r)$, with $\opens\cong\opens(G)$, of the measurement space $\Max C_r^*(G)$ of the reduced C*-algebra of $G$~\cite{msnew}.
\end{example}

\subsection{Change of ``basis''}

There is a way of relating different classical observers that resembles a change of basis in a vector space, as I now explain.
Let $M$ be a measurement space, and let $(\opens_p,r_p)$ and $(\opens_q,r_q)$ be classical observer contexts. Let $X_p$ and $X_q$ be sober spaces such that $\opens_p\cong\topology(X_p)$ and $\opens_q\cong\topology(X_q)$ (these are uniquely defined up to homeomorphism). Write $\closedsets(X_q)$ for the \emph{lower hyperspace} of $X_q$, whose points are the closed sets of $X_q$ and whose topology is the lower Vietoris topology (defined as in section~\ref{Cstar}, now for arbitrary closed sets). The mapping $U\mapsto\ps U$ is an embedding of locales $\topology(X_q)\to\topology(\closedsets(X_q))$.
If we restrict $r_p$ to $\opens_q$ we obtain a join preserving map
$\opens_q\to \opens_p$, and thus also a union preserving map $\topology(X_q)\to\topology(X_p)$. The universal property of the embedding $U\mapsto \ps U$ 
extends this uniquely to a homomorphism of locales $\topology(\closedsets(X_q))\to\topology(X_p)$, which in turn defines a continuous map
\[
\basechg:X_p\to\closedsets(X_q).
\]
This map can be regarded as a topological (``numbers free'') version of change of basis of a vector space: here each ``basis vector'' $x\in X_p$ is a ``combination'' of the ``basis vectors'' contained in $\basechg(x)$.

Let us illustrate this with Schr\"odinger's electron. We have $\opens_x\cong\opens_z\cong\topology(X)$ where $X=\{\vert \spindown\rangle, \vert\spinup\rangle\}$ is a discrete two point space, so
the join preserving map $r_z:\opens_x\to\opens_z$ gives us the inverse image mapping $\basechg^{-1}:\topology(\closedsets(X))\to\topology(X)$:
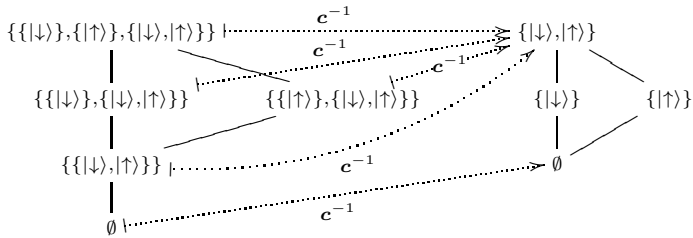
\[\def\objectstyle{\scriptstyle} \def\labelstyle{\scriptstyle}
\xymatrix@-1pc{
\{\{\vert\spindown\rangle\},\{\vert\spinup\rangle\}, \{\vert\spindown\rangle,\vert\spinup\rangle\}\}\ar@{-}[d]\ar@{-}[dr]\ar@{|.>}[rrr]^{\basechg^{-1}}&&&\{\vert \spindown\rangle, \vert\spinup\rangle\}\ar@{-}[d]\ar@{-}[dr]\\
\{\{\vert\spindown\rangle\},\{\vert\spindown\rangle,\vert\spinup\rangle\}\}\ar@{-}[d]\ar@{|.>}[urrr]^{\basechg^{-1}}&\{\{\vert\spinup\rangle\},\{\vert\spindown\rangle,\vert\spinup\rangle\}\}\ar@{-}[dl]\ar@{|.>}[urr]|{\basechg^{-1}}&&\{\vert\spindown\rangle\}\ar@{-}[d]&\{\vert\spinup\rangle\}\ar@{-}[dl]\\
\{\{\vert\spindown\rangle,\vert\spinup\rangle\}\}\ar@{-}[d]\ar@/_20pt/@{|.>}[rrruu]_{\basechg^{-1}}&&&\emptyset\\
\emptyset\ar@{|.>}[rrru]_{\basechg^{-1}}
}
\]
Hence,
$
\basechg(\vert\spindown\rangle)=\basechg(\vert\spinup\rangle)=\{\vert \spindown\rangle, \vert\spinup\rangle\}$.
Regarding $X$ as a placeholder for the bases of eigenvectors of the spin observables $S_x=\frac\hbar 2\sigma_x$ and $S_z=\frac\hbar 2\sigma_z$, $\basechg$ tells us that each eigenvector of $S_z$ is a superposition of all the eigenvectors of $S_x$, but without specifying probability amplitudes: a state either does or does not belong to a superposition.

It can be seen that exactly the same map $\basechg$ would have been obtained if instead of the $x$ direction we had chosen a small angle $\theta\neq 0$ with respect to the $z$ axis. This shows that there is a discontinuity as $\theta$ approaches $0$, for if $\theta=0$ the restriction of $r_z$ is the identity on $\opens_z$, and thus $\basechg(\vert\spindown\rangle)=\{\vert\spindown\rangle\}$ and $\basechg(\vert\spinup\rangle)=\{\vert\spinup\rangle\}$.
In order to remove this discontinuity we would need to consider the complex measures on $X$ that contain the probability amplitudes for each eigenstate, whereas $\basechg$ gives us only the supports of the measures.

\subsection{An example with spin 1}

More illuminating examples can be obtained by considering other spins. For instance, for a spin-1 particle let us again denote by $\boldsymbol z$ the measurement of spin in the $z$ direction with a closed box, whereas the three basic measurements in the same direction, but with the open box, will be denoted by $\boldsymbol z^-$, $\boldsymbol z^0$, and $\boldsymbol z^+$, in correspondence with the three possible recordings on the target.

These measurements can be represented in $\Max M_3(\CC)$ by taking $\boldsymbol z=D_3(\CC)$ to be the maximal abelian subalgebra of diagonal matrices in $M_3(\CC)$, which is generated by the spin observable $S_z$ and the identity matrix $I$, and the three other measurements are one dimensional subspaces spanned by projection matrices:
\[
\boldsymbol z^-=\left(\begin{array}{ccc} 0 & 0 & 0\\
0 & 0 & 0\\
0 & 0 & 1 \end{array}\right)\CC,\quad
\boldsymbol z^0=\left(\begin{array}{ccc} 0 & 0 & 0\\
0 & 1 & 0\\
0 & 0 & 0 \end{array}\right)\CC,\quad
\boldsymbol z^+=\left(\begin{array}{ccc} 1 & 0 & 0\\
0 & 0 & 0\\
0 & 0 & 0 \end{array}\right)\CC\,.
\]
Moreover they are contained in a locale $\opens_z$, namely the following Boolean sublattice of $\Max M_3(\CC)$:
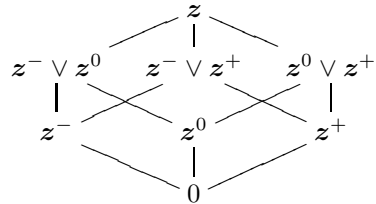
\[
\xymatrix@=10pt{
& {\boldsymbol z}\ar@{-}[dl]\ar@{-}[d]\ar@{-}[dr]\\
{\boldsymbol z^{-}\vee\boldsymbol z^{0}}\ar@{-}[d]\ar@{-}[dr] & {\boldsymbol z^{-}\vee\boldsymbol z^{+}}\ar@{-}[dl]\ar@{-}[dr] & {\boldsymbol z^{0}\vee\boldsymbol z^{+}}\ar@{-}[d]\ar@{-}[dl] \\
{\boldsymbol z^{-}}\ar@{-}[dr] & {\boldsymbol z^{0}}\ar@{-}[d] & {\boldsymbol z^{+}}\ar@{-}[dl] \\
& 0
}
\]

An observer $(\opens_z,r_z)$ can be obtained, similarly to the spin-$\frac 1 2$ case, by defining $r_z$ from the conditional expectation $\Theta_z:M_3(\CC)\to\boldsymbol z$ that restricts matrices to their main diagonal:
\[
\Theta_z(A) = \left(\begin{array}{ccc} a_{11}&0&0\\0&a_{22}&0\\0&0&a_{33}\end{array}\right).
\]

Now let us compare this observer to the analogous one for spin measurements in the $x$ direction. Again we have a locale $\opens_x\subset\Max M_3(\CC)$, analogous to $\opens_z$ but with the $z$ indices replaced by $x$. The measurement $\boldsymbol x$ is generated by the identity matrix and the observable
\[
S_x=\frac\hbar{\sqrt 2}\left(\begin{array}{ccc}0&1&0\\1&0&1\\0&1&0\end{array}\right).
\]
The three measurements of dimension one are spanned by the matrices that project onto the spans of the $x$-spinors
\[
\vert -\rangle_x = \frac 1 2\left(\begin{array}{c}1\\ -\sqrt 2\\1\end{array}\right),\quad
\vert 0\rangle_x = \frac 1 {\sqrt 2}\left(\begin{array}{c}-1\\0\\1\end{array}\right),\quad
\vert +\rangle_x = \frac 1 2\left(\begin{array}{c}1\\ \sqrt 2\\1\end{array}\right),
\]
so we have
\begin{eqnarray*}
\boldsymbol x^-&=&\vert-\rangle_x\langle -\vert_x\CC=\left(\begin{array}{ccc} 1 & -\sqrt 2 & 1\\
-\sqrt 2 & 2 & -\sqrt 2\\
1 & -\sqrt 2 & 1 \end{array}\right)\CC,\\
\boldsymbol x^0&=&\vert 0\rangle_x\langle 0\vert_x\CC=\left(\begin{array}{ccc} 1 & 0 & -1\\
0 & 0 & 0\\
-1 & 0 & 1 \end{array}\right)\CC,\\
\boldsymbol x^+&=&\vert+\rangle_x\langle +\vert_x\CC=\left(\begin{array}{ccc} 1 & \sqrt 2 & 1\\
\sqrt 2 & 2 & \sqrt 2\\
1 & \sqrt 2 & 1 \end{array}\right)\CC.
\end{eqnarray*}

The restriction of $r_z$ to $\opens_x$ yields a mapping which, contrasting with the spin-$\frac 1 2$ examples, does not send all the atoms to the top element, for we have $r_z(\boldsymbol x^{0})=\boldsymbol z^{-}\vee\boldsymbol z^{+}$:
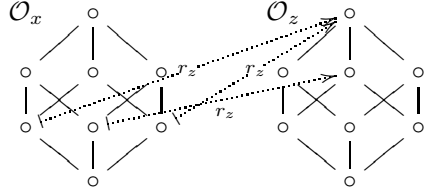
\[
\xymatrix@=10pt{
\opens_x&\circ\ar@{-}[dl]\ar@{-}[d]\ar@{-}[dr]&\hspace*{5mm}&&\opens_z&\circ\ar@{-}[dl]\ar@{-}[d]\ar@{-}[dr]&\hspace*{5mm}\\
\circ\ar@{-}[d]\ar@{-}[rd]&\circ\ar@{-}[dl]\ar@{-}[rd]&\circ\ar@{-}[d]\ar@{-}[dl]
&&\circ\ar@{-}[d]\ar@{-}[rd]&\circ\ar@{-}[dl]\ar@{-}[rd]&\circ\ar@{-}[d]\ar@{-}[dl]\\
\circ\ar@{-}[dr]\ar@{|.>}[rrrrruu]|{r_z}&\circ\ar@{-}[d]\ar@{|.>}[rrrru]_{r_z}&\circ\ar@{-}[dl]\ar@{|.>}[rrruu]|{r_z}&&\circ\ar@{-}[dr]&\circ\ar@{-}[d]&\circ\ar@{-}[dl]\\
&\circ&&&&\circ
}
\]
Hence, identifying both locales with the topology of the discrete space
\[
X=\{\vert -\rangle,\vert 0\rangle,\vert+\rangle\},
\]
the continuous map $\basechg:X\to\closedsets(X)$ such that $\basechg^{-1}\cong r_z$ is given by
\[
\basechg(\vert -\rangle) = \basechg(\vert +\rangle) = X\quad\textrm{and}\quad
\basechg(\vert 0\rangle) = \{\vert -\rangle,\vert+\rangle\}.
\]
Comparing this with the coordinates of the $x$-spinors, again we see that $\basechg$ gives us the supports of the appropriate complex measures on $X$.

\section{Conclusion and discussion}

I have tackled the question of what is meant by a measurement from scratch, in an attempt to provide proof of concept for the idea that measurements are fundamental rather than only relative to arbitrary distinctions between system, apparatus, etc. The reification of this idea into the notion of measurement space  seems to be sufficiently abstract that we do not end up bogged down in concrete details, but also, on the other hand, rich enough that something relevant is conveyed. In particular, despite not being a primitive building block of the theory, a notion of observer has emerged, and there are connections to observer (or context) dependent approaches~\citelist{\cite{Rovelli96}\cite{Doe-Ish-series}\cite{HeLaSp09}\cite{iandurham}\cite{deSilvaBarbosa}\cite{klaas}*{Ch.\ 12}} which deserve being looked at.

Nothing in the definition of measurement space depends on observers, classical or otherwise, which justifies the claim that the latter are ``derived.'' In some sense this makes them less fundamental than measurements, but I remark that this view can be challenged in at least two ways: (1) the definition of classical observer contains more data than just a well behaved subspace of measurements, for the ``glue'' that turns a subspace $\opens$ into a bona fide observer is ``its'' ability, via a retraction onto $\opens$, to approximate more general measurements in a way that conveys the ability of different measurers to communicate the results of their observations; (2)
it is conceivable that further developments may impose consideration of measurement spaces that contain sufficiently rich supplies of observers (just as a spatial locale is a locale with ``enough'' points), thus leading to what should be more appropriately viewed as a joint definition of measurement \emph{and} observer. Rather than being a nuisance, such a situation may in fact offer a consistent mathematical formulation of Bell's ``shifty split,'' interestingly in a way that resonates with Wheeler's~\cite{itfrombit} views regarding the problem of existence.

My purpose in this paper has been to introduce the main definitions, namely of measurement space and classical observer, along with some rationale. Many mathematical details have been glossed over and will appear elsewhere, in particular as regards examples obtained from groupoid C*-algebras (\cf\ Example~\ref{exm:groupoidCstar}). The relation to groupoids, in turn, leads to connections to Schwinger's notion of selective measurement~\cite{Schwinger}, following the insights of Ciaglia et al~\cite{CIM18,CIM19-series}. This should provide welcome hints in any pursuit of a ``picture'' of quantum mechanics based on the ideas of this paper.


\begin{bibdiv}

\begin{biblist}

\bib{AV93}{article}{
  author={Abramsky, Samson},
  author={Vickers, Steven},
  title={Quantales, observational logic and process semantics},
  journal={Math. Structures Comput. Sci.},
  volume={3},
  date={1993},
  number={2},
  pages={161--227},
  issn={0960-1295},
  review={\MR {1224222}},
}

\bib{Araki93}{book}{
  author={Araki, Huzihiro},
  title={Mathematical theory of quantum fields},
  series={International Series of Monographs on Physics},
  volume={101},
  note={Translated from the 1993 Japanese original by Ursula Carow-Watamura; Reprint of the 1999 edition [MR1799198]},
  publisher={Oxford University Press, Oxford},
  date={2009},
  pages={xii+236},
  isbn={978-0-19-956640-2},
  review={\MR {2542202}},
}

\bib{Armstrongetal21}{article}{
  author={Armstrong, Becky},
  author={Clark, Lisa Orloff},
  author={an Huef, Astrid},
  author={Jones, Malcolm},
  author={Lin, Ying-Fen},
  title={Filtering germs: groupoids associated to inverse semigroups},
  journal={Expositiones Mathematicae},
  doi={https://doi.org/10.1016/j.exmath.2021.07.001},
  date={2021},
}

\bib{Ghi-etal03}{article}{
  author={Bassi, Angelo},
  author={Ghirardi, GianCarlo},
  title={Dynamical reduction models},
  journal={Phys. Rep.},
  volume={379},
  date={2003},
  number={5-6},
  pages={257--426},
  issn={0370-1573},
  review={\MR {1979602}},
  doi={10.1016/S0370-1573(03)00103-0},
}

\bib{SGelectrons}{article}{
  author={Batelaan, H.},
  author={Gay, T.J.},
  author={Schwendiman, J.J.},
  title={Stern--Gerlach effect for electron beams},
  journal={Phys. Rev. Lett.},
  volume={79},
  number={23},
  pages={4517--4521},
  date={1997},
}

\bib{Bell90}{article}{
  author={Bell, John},
  title={Against `measurement'},
  journal={Phys. World},
  volume={3},
  date={1990},
  number={8},
  pages={33--40},
  issn={0953-8585},
}

\bib{Bohm-IandII}{article}{
  author={Bohm, David},
  title={A suggested interpretation of the quantum theory in terms of ``hidden'' variables. I and II},
  journal={Physical Rev. (2)},
  volume={85},
  date={1952},
  pages={166--193},
  review={\MR {0046288}},
}

\bib{BLMS87}{article}{
  author={Bombelli, Luca},
  author={Lee, Joohan},
  author={Meyer, David},
  author={Sorkin, Rafael D.},
  title={Space-time as a causal set},
  journal={Phys. Rev. Lett.},
  volume={59},
  date={1987},
  number={5},
  pages={521--524},
  issn={0031-9007},
  review={\MR {899046}},
  doi={10.1103/PhysRevLett.59.521},
}

\bib{CIM18}{article}{
  author={Ciaglia, F. M.},
  author={Ibort, A.},
  author={Marmo, G.},
  title={A gentle introduction to Schwinger's formulation of quantum mechanics: the groupoid picture},
  journal={Modern Phys. Lett. A},
  volume={33},
  date={2018},
  number={20},
  pages={1850122, 8},
  issn={0217-7323},
  review={\MR {3819854}},
  doi={10.1142/S0217732318501225},
}

\bib{CIM19-series}{article}{
  author={Ciaglia, F. M.},
  author={Ibort, A.},
  author={Marmo, G.},
  title={Schwinger's picture of quantum mechanics. I, II, III},
  journal={Int. J. Geom. Methods Mod. Phys.},
  volume={16},
  date={2019},
  pages={1950119\ 31, 1950136\ 32, 1950165\ 37},
}

\bib{CC05}{article}{
  author={Christensen, J. Daniel},
  author={Crane, Louis},
  title={Causal sites as quantum geometry},
  journal={J. Math. Phys.},
  volume={46},
  date={2005},
  number={12},
  pages={122502, 17},
  issn={0022-2488},
  review={\MR {2194023}},
  doi={10.1063/1.2138043},
}

\bib{DeWitt}{article}{
  author={DeWitt, Bryce S.},
  title={Quantum mechanics and reality},
  journal={Physics Today},
  volume={23},
  number={9},
  date={1970},
  pages={30--35},
  doi={10.1063/1.3022331},
}

\bib{Doe-Ish-series}{article}{
  author={D\"{o}ring, A.},
  author={Isham, C. J.},
  title={A topos foundation for theories of physics. I, II, III, IV},
  journal={J. Math. Phys.},
  volume={49},
  date={2008},
  pages={053515 1--25, 053516 1--26, 053517 1--31, 053518 1--29},
}

\bib{iandurham}{article}{
  author={Durham, Ian T.},
  title={An order-theoretic quantification of contextuality},
  journal={Information},
  volume={5},
  date={2014},
  pages={508--525},
  issn={2078-2489},
  doi={10.3390/info5030508},
}

\bib{Everett}{article}{
  author={Everett, Hugh, III},
  title={``Relative state'' formulation of quantum mechanics},
  journal={Rev. Mod. Phys.},
  volume={29},
  date={1957},
  pages={454--462},
  issn={0034-6861},
  review={\MR {0094159}},
  doi={10.1103/revmodphys.29.454},
}

\bib{Ghi-etal90}{article}{
  author={Ghirardi, Gian Carlo},
  author={Pearle, Philip},
  author={Rimini, Alberto},
  title={Markov processes in Hilbert space and continuous spontaneous localization of systems of identical particles},
  journal={Phys. Rev. A (3)},
  volume={42},
  date={1990},
  number={1},
  pages={78--89},
  issn={1050-2947},
  review={\MR {1061931}},
  doi={10.1103/PhysRevA.42.78},
}

\bib{bigdomainbook}{book}{
  author={Gierz, G.},
  author={Hofmann, K. H.},
  author={Keimel, K.},
  author={Lawson, J. D.},
  author={Mislove, M.},
  author={Scott, D. S.},
  title={Continuous lattices and domains},
  series={Encyclopedia of Mathematics and its Applications},
  volume={93},
  publisher={Cambridge University Press, Cambridge},
  date={2003},
  pages={xxxvi+591},
  isbn={0-521-80338-1},
  review={\MR {1975381}},
}

\bib{HeLaSp09}{article}{
  author={Heunen, Chris},
  author={Landsman, Nicolaas P.},
  author={Spitters, Bas},
  title={A topos for algebraic quantum theory},
  journal={Comm. Math. Phys.},
  volume={291},
  date={2009},
  number={1},
  pages={63--110},
  issn={0010-3616},
  review={\MR {2530156}},
  doi={10.1007/s00220-009-0865-6},
}

\bib{pointless}{article}{
  author={Johnstone, Peter T.},
  title={The point of pointless topology},
  journal={Bull. Amer. Math. Soc. (N.S.)},
  volume={8},
  date={1983},
  number={1},
  pages={41--53},
  issn={0273-0979},
  review={\MR {682820 (84f:01043)}},
}

\bib{decoherencebook}{book}{
  author={Joos, E.},
  author={Zeh, H. D.},
  author={Kiefer, C.},
  author={Giulini, D.},
  author={Kupsch, J.},
  author={Stamatescu, I.-O.},
  title={Decoherence and the appearance of a classical world in quantum theory},
  edition={2},
  publisher={Springer-Verlag, Berlin},
  date={2003},
  pages={xii+496},
  isbn={3-540-00390-8},
  review={\MR {2148270}},
  doi={10.1007/978-3-662-05328-7},
}

\bib{klaas}{book}{
  author={Landsman, Klaas},
  title={Foundations of quantum theory},
  series={Fundamental Theories of Physics},
  volume={188},
  subtitle={From classical concepts to operator algebras},
  publisher={Springer, Cham},
  date={2017},
  pages={xv+881},
  isbn={978-3-319-51776-6},
  isbn={978-3-319-51777-3},
  review={\MR {3643288}},
  doi={10.1007/978-3-319-51777-3},
}

\bib{M86}{article}{
  author={Mulvey, Christopher J.},
  title={\&},
  note={Second topology conference (Taormina, 1984)},
  journal={Rend. Circ. Mat. Palermo (2) Suppl.},
  number={12},
  date={1986},
  pages={99--104},
  review={\MR {853151 (87j:81017)}},
}

\bib{Curacao}{misc}{
  author={Mulvey, Christopher J.},
  title={Quantales},
  note={Invited talk at the Summer Conference on Locales and Topological Groups (Cura\c {c}ao, 1989)},
}

\bib{MP1}{article}{
  author={Mulvey, Christopher J.},
  author={Pelletier, Joan Wick},
  title={On the quantisation of points},
  journal={J. Pure Appl. Algebra},
  volume={159},
  date={2001},
  number={2-3},
  pages={231--295},
  issn={0022-4049},
  review={\MR {1828940 (2002g:46126)}},
}

\bib{Penrose2014}{article}{
  author={Penrose, Roger},
  title={On the gravitization of quantum mechanics 1: Quantum state reduction},
  journal={Found. Phys.},
  volume={44},
  date={2014},
  number={5},
  pages={557--575},
  issn={0015-9018},
  review={\MR {3210210}},
  doi={10.1007/s10701-013-9770-0},
}

\bib{Re01}{article}{
  author={Resende, Pedro},
  title={Quantales, finite observations and strong bisimulation},
  journal={Theoret. Comput. Sci.},
  volume={254},
  date={2001},
  number={1-2},
  pages={95--149},
  issn={0304-3975},
  review={\MR {1816827}},
}

\bib{Re07}{article}{
  author={Resende, Pedro},
  title={\'Etale groupoids and their quantales},
  journal={Adv. Math.},
  volume={208},
  date={2007},
  number={1},
  pages={147--209},
  issn={0001-8708},
  review={\MR {2304314 (2008c:22002)}},
}

\bib{GSQS}{article}{
  author={Resende, Pedro},
  title={Groupoid sheaves as quantale sheaves},
  journal={J. Pure Appl. Algebra},
  volume={216},
  date={2012},
  number={1},
  pages={41--70},
  issn={0022-4049},
  review={\MR {2826418}},
  doi={10.1016/j.jpaa.2011.05.002},
}

\bib{QFB}{article}{
  author={Resende, Pedro},
  title={Quantales and Fell bundles},
  journal={Adv. Math.},
  volume={325},
  date={2018},
  pages={312--374},
  issn={0001-8708},
  review={\MR {3742593}},
  doi={10.1016/j.aim.2017.12.001},
}

\bib{SGQ}{article}{
  author={Resende, Pedro},
  title={The many groupoids of a stably Gelfand quantale},
  journal={J. Algebra},
  volume={498},
  date={2018},
  pages={197--210},
  doi={10.1016/j.jalgebra.2017.11.042},
}

\bib{msnew}{article}{
  author={Resende, Pedro},
  title={On the geometry of physical measurements: topological and algebraic aspects},
  eprint={https://arxiv.org/abs/2005.00933v3},
  date={2021},
}

\bib{RS2}{article}{
  author={Resende, Pedro},
  author={Santos, Jo\~{a}o Paulo},
  title={Linear structures on locales},
  journal={Theory Appl. Categ.},
  volume={31},
  date={2016},
  pages={Paper No. 20, 502--541},
  issn={1201-561X},
  review={\MR {3513965}},
  eprint={http://www.tac.mta.ca/tac/volumes/31/20/31-20.pdf},
}

\bib{Rosenthal1}{book}{
  author={Rosenthal, Kimmo I.},
  title={Quantales and Their Applications},
  series={Pitman Research Notes in Mathematics Series},
  volume={234},
  publisher={Longman Scientific \& Technical},
  place={Harlow},
  date={1990},
  pages={x+165},
  isbn={0-582-06423-6},
  review={\MR {1088258 (92e:06028)}},
}

\bib{Rovelli96}{article}{
  author={Rovelli, Carlo},
  title={Relational quantum mechanics},
  journal={Internat. J. Theoret. Phys.},
  volume={35},
  date={1996},
  number={8},
  pages={1637--1678},
  issn={0020-7748},
  review={\MR {1409502}},
  doi={10.1007/BF02302261},
}

\bib{Schwinger}{article}{
  author={Schwinger, Julian},
  title={The algebra of microscopic measurement},
  journal={Proc. Nat. Acad. Sci. U.S.A.},
  volume={45},
  date={1959},
  pages={1542--1553},
  issn={0027-8424},
  review={\MR {112598}},
  doi={10.1073/pnas.45.10.1542},
}

\bib{deSilvaBarbosa}{article}{
  author={de Silva, Nadish},
  author={Barbosa, Rui Soares},
  title={Contextuality and noncommutative geometry in quantum mechanics},
  journal={Comm. Math. Phys.},
  volume={365},
  date={2019},
  number={2},
  pages={375--429},
  issn={0010-3616},
  review={\MR {3907949}},
  doi={10.1007/s00220-018-3222-9},
}

\bib{Smolin06}{article}{
  author={Smolin, Lee},
  title={The case for background independence},
  conference={ title={The structural foundations of quantum gravity}, },
  book={ publisher={Oxford Univ. Press, Oxford}, },
  date={2006},
  pages={196--239},
  review={\MR {2387728}},
  doi={10.1093/acprof:oso/9780199269693.003.0007},
}

\bib{Stoy}{book}{
  author={Stoy, Joseph E.},
  title={Denotational semantics: the Scott-Strachey approach to programming language theory},
  series={MIT Press Series in Computer Science},
  volume={1},
  note={Reprint of the 1977 original; With a foreword by Dana S. Scott},
  publisher={MIT Press, Cambridge, Mass.-London},
  date={1981},
  pages={xxx+414},
  isbn={0-262-69076-4},
  review={\MR {629830}},
}

\bib{topologyvialogic}{book}{
  author={Vickers, Steven},
  title={Topology via logic},
  series={Cambridge Tracts in Theoretical Computer Science},
  volume={5},
  publisher={Cambridge University Press, Cambridge},
  date={1989},
  pages={xvi+200},
  isbn={0-521-36062-5},
  review={\MR {1002193}},
}

\bib{Vi07}{article}{
  author={Vickers, Steven},
  title={Locales and toposes as spaces},
  book={ title={Handbook of spatial logics} editor={M.\ Aiello}, editor={I.\ Pratt-Hartmann}, editor={J.\ Van Benthem}, publisher={Springer, Dordrecht}, },
  date={2007},
  pages={429--496},
  review={\MR {2393892}},
}

\bib{Wheeler}{article}{
  author={Wheeler, John A.},
  title={Assessment of Everett's ``relative state'' formulation of quantum theory},
  journal={Rev. Mod. Phys.},
  volume={29},
  date={1957},
  pages={463--465},
  issn={0034-6861},
  review={\MR {0094160}},
  doi={10.1103/revmodphys.29.463},
}

\bib{itfrombit}{article}{
  author={Wheeler, John Archibald},
  title={Information, physics, quantum: the search for links},
  conference={ title={Foundations of quantum mechanics in the light of new technology}, address={Tokyo}, date={1989}, },
  book={ publisher={Phys. Soc. Japan, Tokyo}, },
  date={1990},
  pages={354--368},
  review={\MR {1105973}},
}

\bib{Zeh70}{article}{
  author={Zeh, Heinz-Dieter},
  title={On the interpretation of measurement in quantum theory},
  journal={Found. Phys.},
  volume={1},
  date={1970},
  number={1},
  pages={69--76},
}

\bib{Zeh73}{article}{
  author={Zeh, Heinz-Dieter},
  title={Towards a quantum theory of observation},
  journal={Found. Phys.},
  volume={3},
  date={1973},
  number={1},
  pages={109--116},
}

\bib{Zurek03}{article}{
  author={Zurek, Wojciech Hubert},
  title={Decoherence, einselection, and the quantum origins of the classical},
  journal={Rev. Modern Phys.},
  volume={75},
  date={2003},
  number={3},
  pages={715--775},
  issn={0034-6861},
  review={\MR {2037624}},
  doi={10.1103/RevModPhys.75.715},
}

\end{biblist}

\end{bibdiv}
\end{document}